\documentclass[prl,twocolumn,showpacs]{revtex4}

\usepackage{epsfig}

\begin{document}

\title{Surface Driven Bulk Reconstruction of Gold Nanorods}

\author{Yanting Wang,$^a$ S. Teitel$^a$ and Christoph Dellago$^b$}

\affiliation{$^a$Department of Physics and Astronomy, University of
Rochester, Rochester, NY 14627\\
$^b$Institute for Experimental Physics, University of
Vienna, Boltzmanngasse 5, 1090 Vienna, Austria}

%-------------------------------------------------------------------------
%  ABSTRACT
%-------------------------------------------------------------------------
\vspace*{1cm}
\begin{abstract}
Molecular dynamic simulations are used to study the heating of a  gold 
nanorod of $2624$ atoms. We show that roughening of surface $\{110\}$
facets leads to a shape transformation and structural rearrangement of 
surface and bulk atoms in the rod, in order to reach a more stable
configuration.  Our results show that the stability of the nanorod is governed
by the free energetics of the surface facets. 
\end{abstract}
\pacs{64.70.Nd, 61.46.+w, 82.60.Qr}
\maketitle

%------------------------------------------------------------------------
% BODY
%------------------------------------------------------------------------

The properties of metallic clusters of various shapes, on the nanometer length scale,
can be dramatically different from the corresponding bulk material
due to their very large surface-to-volume ratio.  The unique optical
and mechanical properties of such nanoclusters hold great promise
for the development of nanotechnology.  Gold nanoclusters, for example,
have found application in such diverse fields as
nano-lithography \cite{R1}, catalysis \cite{R2},
nano-bioelctronic devices \cite{R3}, and ion detection \cite{R4}.
It is therefore of great interest to study the  behavior and stability
of such nano-structures.

In this work we consider gold nanorods of low aspect
ratio $\sim 3$.  Using molecular dynamic (MD) simulations of a rod with
a few thousand atoms, we find that the stability of the nanorod
upon heating is governed
primarily by the energetics of it's surface.  When we start with a rod
which has large $\{110\}$ surface facets, the roughening of
these facets upon heating leads to a shape transformation of the
rod to a shorter and wider structure.   The surface reconstructs to
form higher stability $\{111\}$ facets, while the fcc interior completely
reorients to align with the new facet planes.  When we start with a
rod which has predominantly $\{111\}$ facets on its surface, the
rod remains stable until melting.

\begin{figure}
\epsfxsize=7truecm
\epsfbox{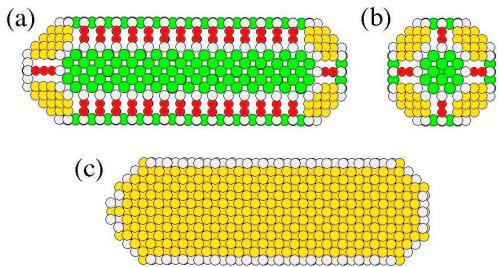}
\caption{Initial configuration of the gold nanorod with $N=2624$ atoms
and aspect ratio $3$: (a) side view, (b)
end view down the long axis, (c) cross-sectional view parallel to long axis.
In (a) and (b), yellow atoms represent $\{111\}$ facets, green atoms $\{100\}$
facets, and red atoms $\{110\}$ facets; white atoms are on the edges.  
In (c), yellow atoms are those with a
local fcc structure, white atoms are on the surface; the cross-sectional view
in (c) shows atoms in the square arrangement of a $\{100\}$ plane
of the fcc lattice.
}
\label{f1}
\end{figure}

The initial rod configuration that we consider, shown in Fig.\,\ref{f1},  is one which
has been proposed  \cite{Wang} to apply to recent experiments \cite{Chang,Link1,Link2}
on the laser heating of gold nanorods.  
The interior of the rod is a pure fcc lattice.  The surface of the rod
consists of four large $\{100\}$ and four large $\{110\}$ facets oriented
parallel to the rod axis.  The ends of the rod have a $\{001\}$ facet and
four small $\{111\}$ facets connecting the $\{110\}$ and the $\{001\}$ facets.
These experiments found that,
upon heating, such rods underwent a shape transformation to bent, twisted, shorter,
wider, and $\phi$-shaped clusters.  Transmission electron miscroscopy studies \cite{Link2}
observed point and planar internal defects to accompany such shape transformations.
Recent MD simulations \cite{Wang_Dellago}
of such rods, using a continuous heating procedure meant to model the laser heating
of experiments, found  similar shape transformations.  These simulations found
the shape transformation to be accompanied by a structural 
change in which planes of interior atoms shift, converting local fcc structure to hcp.
The extent and stability of these interior rearrangements was found to depend upon both
the heating rate and the number of atoms in the cluster, but no
specific mechanism or energetic argument for this structural rearrangement was proposed.
In this paper we present new simulations carried out with
a much slower ``quasi-equilibrium" heating that allows
the rod more time to approach its lowest free energy configuration.
Our results make it clear that it is the energetics of the surface that is driving
the shape and structural transformation.
%  A somewhat similar mechanism, involving
%instabilities of the $\{110\}$ surface, but otherwise different in detail, has
%been proposed earlier based on  experimental observations \cite{Link2,Wang2}.

We use the empirical ``glue" 
potential \cite{glue} to model the many body interactions of the gold atoms in our
simulated nanorod, and we integrate 
the classical equations of motion for the atoms using the velocity
Verlet algorithm \cite{verlet} with a time step of $4.3$ fs.  However instead of 
increasing the kinetic energy at each MD step to model
continuous heating as in Ref.\,[\onlinecite{Wang_Dellago}], we now use
the Gaussian isokinetic thermostat \cite{isokinetic} to keep the total kinetic energy fixed 
at a constant temperature $T$; after each MD step, all velocities are rescaled
by a constant factor so as to keep $\langle (1/2)mv^2\rangle=(3/2)k_BT$ fixed.
Our procedure conserves total linear and
angular momentum, which are set to zero, so that our rod does not drift
or rotate throughout our simulation.
At each fixed $T$ we carry out
$10^7$ MD steps, for a simulated time of $43$ ns, before increasing 
the temperature in jumps of $100$ K.  Our effective heating rate is therefore
$\sim 2.3\times 10^{9}$ K/s, more than three orders of magnitude slower
than the continuous heating rates of $2-7\times 10^{12}$ K/s used in Ref.\,[\onlinecite{Wang_Dellago}].
We use a rod of $N=2624$ atoms with initial aspect ratio of $3$, as shown in
Fig.\,\ref{f1}.
The length of the rod, parallel to its long axis, is $7.38$ nm and its crossectional area
has a diameter of $2.46$ nm.
We do a short equilibration for $430$ ps ($10^5$ MD steps) at $5$ K in order
to relax the surface atoms of the rod from their initial fcc positions, 
before starting to heat the rod.

\begin{figure}
\epsfxsize=8.truecm
\epsfbox{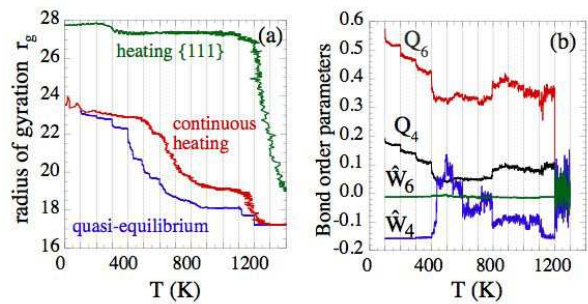}
\caption{(a) Radius of gyration $r_g$ vs. temperature $T$ for
the rod of Fig.\,\ref{f1}: quasi-equlibrium heating (blue) compared
to the continuous heating of Ref.\,[\protect\onlinecite{Wang_Dellago}]
(red); continuous heating of the rod with structure of Fig.\,\protect\ref{f4} (green).  
(b) Bond order parameters, $Q_4$, $Q_6$, $\hat W_4$
and $\hat W_6$, averaged over all atoms internal to the rod, vs. temperature $T$,
for quasi-equlibrium heating of the rod of Fig.\,\ref{f1}.
}
\label{f2-3}
\end{figure}

As a signature of the shape change of our nanocluster we measure the
radius of gyration, $r_g$, defined by
$r_g^2=(1/N)\sum_i\left|{\bf r}_i-{\bf r}_c\right|^2$,
where ${\bf r}_i$ is the position of atom $i$ and ${\bf r}_c$ is
the center of mass.  In Fig.\,\ref{f2-3}a we plot our results for $r_g$
as the system is heated; the blue curve is for our above ``quasi-equilibrium"
heating.   The vertical dotted lines separate bins of constant temperature
simulation, where the temperature is equal to the value at the left end point
of the bin; the data plotted within each bin represents the instantaneous value
of $r_g$ as a function of increasing time at the constant temperature.
At the end of each bin the temperature is increased by a jump of $100$ K.
We plot our data this way, instead of as an average value at each $T$, 
to highlight that significant shape relaxation occurs even at
constant $T$.  For comparison, we plot $r_g$ for the continous heating
(red curve) of Ref.\,[\onlinecite{Wang_Dellago}] for the heating rate of
$7\times 10^{12}$ K/s.  We see that the curves are qualitatively similar,
with the onset of a plateau around $800$ K, however the present
quasi-equilibrium heating allows the rod to relax to smaller $r_g$ values, before
the rod melts at $T\sim 1200$ K.
The decrease in the radius of gyration reflects the shape
transformation to a shorter wider rod of smaller aspect ratio.

To investigate the local structure within the cluster, we use the method
of bond orientational order parameters \cite{bop}.  These parameters measure the
orientation of bonds connecting a given atom to its nearest neighbors, and
provide a convenient means of determining the local crystalline structure
of an atom.  In particular, we measure the $6$-fold and $4$-fold orientation
parameters, $Q_6$, $\hat W_6$, $Q_4$, and $\hat W_4$.  We refer the
reader to the literature for their definitions \cite{Wang_Dellago,bop}, 
and in Table\,\ref{t1} we give their
values for several periodic three dimensional crystal structures.  We also,
in Table I, give the values of these parameters as computed for atoms on particular
low index planar surfaces of an fcc bulk crystal; for these two dimensional
parameters we average only over bonds connecting an atom with its neighbors
in the specified plane.  In Fig.\,\ref{f2-3}b we plot these order parameters for our
rod, averaging over only atoms internal to the rod (i.e. we exclude surface atoms since
these have fewer nearest neighbor bonds).  As for $r_g$, we plot our data as the
instantaneous value as a function of increasing simulation time, for bins of
constant temperature (indicated by the dotted vertical lines).

\begin{table}
\caption{\label{t1}Bond order parameters for face-centered-cubic (fcc),
hexagonal close-packed (hcp), simple cubic (sc), body-centered-cubic (bcc),
and low index fcc planes.}
\begin{ruledtabular}
\begin{tabular}{lccrr}
Geometry & $Q_4$ & $Q_6$ & $\hat{W}_4$ & $\hat{W}_6$
\\ \hline
fcc & 0.190\,94 & 0.574\,52 & $-$0.159\,32 & $-$0.013\,16
\\
hcp & 0.097\,22 & 0.484\,76 & 0.134\,10 & $-$0.012\,44
\\
sc  & 0.763\,76 & 0.353\,55 & 0.159\,32 & 0.013\,16
\\
bcc & 0.082\,02 & 0.500\,83 & 0.159\,32 & 0.013\,16
\\
\{110\} & 1 & 1 & 0.134\,10 & $-$0.093\,06
\\
\{100\} & 0.829\,16 & 0.586\,30 & 0.124\,97 & $-$0.007\,21
\\
\{111\} & 0.375\,00 & 0.740\,83 & 0.134\,10 & $-$0.046\,26
\\ 
\end{tabular}
\end{ruledtabular}
\end{table}

Comparing with Table\,\ref{t1}, we see that the rod maintains its fcc
structure until about $400$ K.  Then, from around $400$ K to about
$800$ K, there is a rise to positive values in $\hat W_4$, and a decrease
in $Q_6$ and $\hat W_6$, suggestive of a more hcp-like structure.
Above $800$ K, the values return to their fcc-like values.

\begin{figure}
\epsfxsize=6.5truecm
\epsfbox{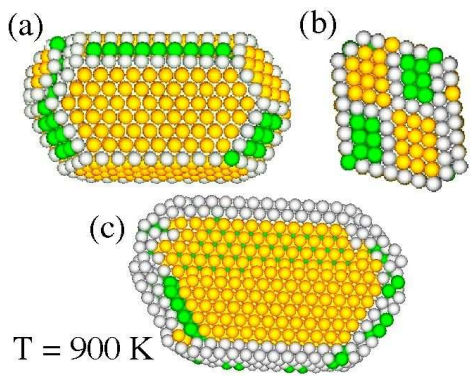}
\caption{Configuration of the nanorod after quasi-equilibrium heating
to $900$ K.  (a) side view, and (b) end view down the long axis, after
peeling away the surface and the first sub-surface layer; 
yellow atoms are $\{111\}$ facets, green atoms are 
$\{100\}$ facets, and white are edge atoms.  (c) cross-sectional view parallel
to the long axis; yellow atoms have a local fcc structure, green atoms have a
local hcp structure, and white atoms are neither.  The cross-sectional view in
(c) shows atoms in the close-packed hexagonal arrangement of a
$\{111\}$ plane of the fcc lattice.
}
\label{f4}
\end{figure}

We now focus on the structure of the rod in the high temperature 
plateau region where $r_g$ stabilizes to a constant.
In Fig.\,\ref{f4} we show the configuration of the rod at $T=900$ K,
in the constant plateau region before melting.  The views of the
rod shown in Figs.\,\ref{f4}a,b,c are the same orientations as shown
for the initial configuration in Figs.\,\ref{f1}a,b,c.  In order to better
illustrate the order of the rod, we first pick an instantaneous
configuration sampled from the middle of the $T=900$ K simulation,
and use the conjugate gradient method \cite{conjgrad} to quench local thermal fluctuations.
At such high temperatures, the surface can be partially disordered
compared to the interior, due to the diffusion of atoms on and near
facet edges and vertices \cite{WTD1,WTD2}.  We therefore use the 
{\it cone algorithm} \cite{WTD2}
to identify and peel away atoms on the surface and in the first sub layer
below it, and in Figs.\,\ref{f4}a,b show the configuration of the second
sub layer of the rod.  We see a very regular shape covered almost
completely with stable $\{111\}$ facets.  Based on the values in Table\,\ref{t1},
we use the following criteria to identify atoms in this layer as belonging to
particular low index planes: $\{111\}$ if $0.7<Q_6<0.9$ and $-0.08<\hat W_6<-0.02$;
$\{100\}$ if $Q_6<0.7$ and $\hat W_6>-0.02$; and $\{110\}$ if
$Q_6>0.9$ and $\hat W_6<-0.08$.  Atoms in Figs.\,\ref{f4}a,b have been
colored accordingly.

The cross-sectional view in
Fig.\,\ref{f4}c shows an almost pure fcc interior, as was the case
for the initial configuration, however we now see
a close packed hexagonal structure characteristic of a $\{111\}$ plane
of the fcc lattice, rather than the $\{100\}$ plane seen in the cross-sectional
view of Fig.\,\ref{f1}c.  We thus see one of our main results:
in order to align with the new $\{111\}$
surface facets, the bulk fcc structure has completely reconstructed
itself to a new orientation.  For interior atoms, we use the following criteria
to identify the local crystal structure: fcc if $Q_4>0.17$ and $\hat W_4<-0.10$;
hcp if $Q_4<0.13$ and $\hat W_4>0.07$.  Atoms in Figs.\,\ref{f4}c have been
colored accordingly.

\begin{figure}
\epsfxsize=7.5truecm
\epsfbox{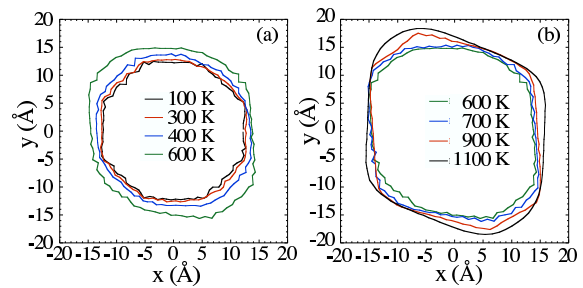}
\caption{Average cross-sectional shape, viewed down the long axis of the rod,
for different temperatures.
}
\label{f5}
\end{figure}

To see how the rod evolves from its initial configuration (Fig.\,\ref{f1}) to
its reconstructed shape (Fig.\,\ref{f4}), we consider the average cross-sectional
shape in a plane transverse to the long axis of the rod.  We compute this average 
shape as follows.  For each instantaneous configuration
we first eliminate all atoms on the end caps of the rods, and all interior atoms of the
rod, and then project the
remaining surface atoms into the $xy$ plane, perpendicular to the long axis of the rod.
Placing the origin at the resulting center of mass, we divide the plane into
$100$ equal polar angles, and then compute the average position of all surface atoms in
each angular division.  This result is then averaged over $1000$ different 
instantaneous configurations
sampled uniformly throughout the simulated time of $43$ ns at each temperature $T$.
We plot the resulting average cross-sectional shapes, for several different $T$, in
Fig.\,\ref{f5}.

At low $T$ we see the octagonal cross-section of the initially constructed rod
of Fig.\,\ref{f1}, with the flat edges representing the initial $\{100\}$ and $\{110\}$
facets.  The shape stays roughly the same until about $400$ K.
Somewhere between $300-400$ K, the shape becomes rounder and the initial
flat edges disappear.  As $T$ increases further, the cross-sectional area grows, 
representing the shape transformation to a shorter and wider rod of lower aspect ratio,
and we see new flat facets develop and grow in new directions.  At $900$ K we see the
fully faceted shape shown in Fig.\,\ref{f4}.

The disappearance of the initial flat facets, with the 
resulting rounding of the average shape, is a signature of the roughening transition
of those surfaces \cite{roughen}.  For {\it macroscopic} gold samples, it is known from
experiments that the $\{110\}$ surface roughens at $680$ K \cite{s110}, while the
$\{100\}$ surface disorders at $\sim 1170$ K \cite{s100}, below the bulk melting temperature
of $1337$ K.  In contrast, the $\{111\}$ surface is believed to remain stable
up to, and even above, the bulk melting \cite{s111}.
The initiation of the shape change that we find in our nanorod at $400$ K
is most likely a consequence of the roughening transition of the $\{110\}$ facets,
which has been shifted to lower temperature due
to large finite size effects in our relatively small rod (just as the bulk melting
transition can be greatly reduced by finite size effects \cite{Buffat}).
After this roughening, the surface reconstructs to form mostly
lower free energy $\{111\}$ facets, which remain stable until melting.
Note that the fully faceted cross-sectional shape at $900$ K contains four 
large sides and two short sides; the former are the $\{111\}$ facets, while
the latter are $\{100\}$ facets.  By $1100$ K, these $\{100\}$ facets have
been replaced by a smoothly curved surface.  We infer that this is due to the
disordering transition of the $\{100\}$ surface, reduced somewhat in temperature
due to finite size effects.

\begin{figure}
\epsfxsize=7.6truecm
\epsfbox{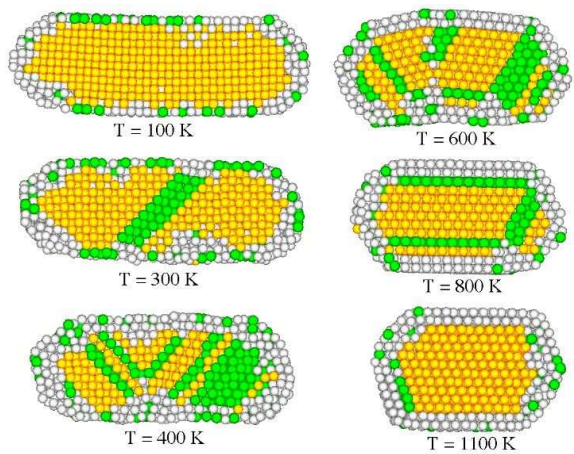}
\caption{Cross-sectional view for various temperatures.  Atoms are colored
according to local crystal structure: fcc is yellow, hcp is green, neither is white.
}
\label{f6}
\end{figure}

In order to see how the interior fcc structure of the rod reconstructs itself
to a new orientation, we show in Fig.\,\ref{f6} cross-sectional views of the
rod at various temperatures.  We color the atoms according to their local
crystal structure, using the criteria given above: fcc is yellow, hcp is green,
neither is white.  Initially, the interior is pure fcc, oriented so that the cross-sectional
view shows a $\{100\}$ plane of atoms.  As temperature increases, we see that 
the shape and structural transformation is
accompanied by the appearance of hcp planes inside the rod interior, due
to the sliding of $\{111\}$ planes.  As temperature further increases, the surface
becomes less ordered, and more $\{111\}$ planes with different orientation slide.
Around $800$ K,  the surface has reordered and the interior fcc lattice has
reoriented so that the cross-sectional view now shows a predominantly $\{111\}$
plane of atoms.  At $1100$ K, the interior has completely reordered to pure fcc,
but with the new orientation.

Such behavior as described above may well exist in other
simple elemental metals.  We note that many such metals similarly have a
roughening transition $T_R$ for the $\{110\}$ surface that is significantly below
the bulk melting $T_m$.   Silver, for example, has $T_R=600$ K and $T_m=1235$ K, 
with a similar ratio of $T_R/T_m$ as gold \cite{silver}.
Lead has $T_R=415$ K and
$T_m=601$ K, for a somewhat larger $T_R/T_m$ than gold \cite{lead}.

%To conclude, we find that, upon very slow heating, a gold nanorod 
%with large initial $\{110\}$ and $\{100\}$
%facets undergoes a shape transformation initiated by the roughening of the $\{110\}$
%surface.  After this roughening, the surface of the rod
%reconstructs to form stable $\{111\}$
%facets, while the fcc interior of the rod reconstructs to a new orientation so as to
%be compatible with the new $\{111\}$ surface facets.  The result is a perfect fcc rod,
%of lower aspect ratio than initially, covered almost entirely by stable $\{111\}$ facets.
%The mechanism for the reorientation of the interior is the sliding of close-packed
%$\{111\}$ planes interior to the rod.  Although our simulated rod of $2624$ atoms
%is much smaller than those in experiments, which may have $~10^6$ atoms, 
%the general mechanism outlined above may still have
%relevance for the observed shape transformations and stability of gold nanorods
%in experimentally studied systems.

Finally, in order to verify that the roughening of the $\{110\}$ facets,
rather than just the minimization of total surface area, is indeed the mechanism
for the shape transformation, we study
the stability of a gold nanorod with an aspect ratio of $\sim 3$, but with
an initial structure similar to that of Fig.\,\ref{f4}, with a surface predominantly
covered by $\{111\}$ facets.  We use a continuous heating MD simulation with a
heating rate of $7\times 10^{12}$ K/s to model laser heating experiments, for a
rod with $3411$ atoms.  
Our results for $r_g$ vs $T$ are shown in Fig.\,\ref{f2-3}a (green curve).  Unlike
the intial rod of Fig.\,\ref{f1}, we now find that the rod remains
stable, with no significant shape or structural rearrangement, up until the rod melting 
temperature.  We conclude that the stability of gold, and presumably other metallic,
nanorods is crucially dependent upon the structure of the rod surface.

This work was funded in part by DOE grant DE-FG02-89ER14017.
%-------------------------------------------------------------------------
%  REFERENCES
%-------------------------------------------------------------------------

%

\end{document}